\documentclass[12pt,preprint]{aastex}

\slugcomment{Submitted: ApJ, March 03, 2013; Revised: August 09, 2013}

\shorttitle{RR~Lyrae Mid-IR PL Relation}
\shortauthors{Madore et al.}

\begin{document}


\title{\Large A Preliminary Calibration of the\\RR Lyrae Period-Luminosity Relation at\\Mid-Infrared Wavelengths: WISE Data}

\medskip
\medskip
\medskip
\medskip
\medskip

\author{\bf }
\author{\bf }
\author{\large\bf Barry F. Madore}
\affil{Observatories of the Carnegie
  Institution of Washington \\ 813 Santa Barbara St., Pasadena, CA
  ~~91101} 

\author{\large\bf Douglas Hoffman }
\affil{Infrared Processing and Analysis Center
  \\ 770 South Wilson, Pasadena, CA  ~~91125}

\author{\large\bf Wendy L. Freedman, Juna A. Kollmeier, Andy Monson }

\author{\large\bf S. Eric Persson, Jeff A. Rich Jr., Victoria Scowcroft, Mark Seibert}
\affil{Observatories of the Carnegie
  Institution of Washington \\ 813 Santa Barbara St., Pasadena, CA
  ~~91101} 
\medskip
\medskip

\email{barry@obs.carnegiescience.edu, dhoffman@ipac.caltech.edu\\
wendy@obs.carnegiescience.edu, jak@obs.carnegiescience.edu \\ 
amonson@obs.carnegiescience.edu, persson@obs.carnegiescience.edu \\
jrich@obs.carnegiescience.edu, vs@obs.carnegiescience.edu, 
mseibert@obs.carnegiescience.edu}

\author{\bf }
\author{\bf }
\author{\bf }
\author{\bf }
\author{\bf }
\author{\bf }


\vfill\eject
\medskip
\medskip
\medskip
\begin{abstract}
Using time-resolved, mid-infrared data from WISE and geometric
parallaxes from {\it HST} for four Galactic RR Lyrae variables, we
derive the following Population II Period-Luminosity (PL) relations
for the WISE[W1], [W2] and [W3] bands at 3.4, 4.6 \& 12~$\mu$m,
respectively:
$$ M_{[W1]} = -2.44 ~(\pm 0.95) \times log(P) -1.26 ~(\pm0.25)~~ \sigma = 0.10$$
$$ M_{[W2]} = -2.55 ~(\pm 0.89) \times log(P) -1.29 ~(\pm0.23)~~ \sigma = 0.10$$
$$ M_{[W3]} = -2.58 ~(\pm 0.97) \times log(P) -1.32 ~(\pm0.25)~~ \sigma = 0.10$$ 
The slopes and the scatter around the fits are consistent with a smooth
extrapolation of those same quantities from previously-published K-band
observations at 2.2~$\mu$m, where the asymptotic (long-wavelength)
behavior is consistent with a Period-Radius relation having a slope of
0.5. No obvious correlation with metallicity (spanning 0.4~dex in
[Fe/H]) is found in the residuals of the four calibrating RR Lyrae
stars about the mean PL regression line.

\end{abstract}


\vfill\eject
\medskip
\medskip
\medskip
\medskip
\medskip
\medskip
\medskip
\medskip
\medskip
\vfill\eject
\section{Introduction}

The advantages of moving the Population I calibration of the Classical
Cepheid Period-Luminosity relation (the Leavitt Law) from the optical
to the infrared were outlined some three decades ago by McGonegal et
al. (1983), and they have been borne out repeatedly over the years, as
reviewed and elaborated upon by Freedman \& Madore (2010). But only
now is it possible to extend these same advantages to the parallel
path offered by the Population II variable stars (the short-period
RR~Lyrae variables and their longer-period siblings the W~Virginis
stars). Two impressive accomplishments have made this possible: (1)
the completion of the WISE mission (Wright et al. 2010) and the
release of its sky survey of point sources measured in the
mid-infrared in four bands ranging from 3.4 to 22~$\mu$m, and then (2)
the innovative application of the {\it HST} Fine-Guidance Sensor ({\it
  FGS}) ~cameras to the determination of trigonometric parallaxes to
four field RR Lyrae variables by Benedict et al. (2011).

The many, now well-known, advantages of calibrating and using
period-luminosity relations in the mid-IR include the following: (1)
the effects of line-of-sight extinction are reduced with respect to
optical observations by at least an order of magnitude for even the
shortest wavelength ([W1] at 3.4~$\mu$m) observations, (2) concerns
about the systematic impact of the possible non-universality of the
reddening law are similarly reduced by going away from the optical and
into the mid-IR, (3) the total amplitude of the light variation of the
target star during its pulsation cycle is greatly reduced because of
the largely diminished contribution of temperature variation to the
change in surface brightness, in comparison to the much smaller (but
essentially irreducible) wavelength-independent radius/areal
variations, (4) the corresponding collapse in the width (i.e.,
intrinsic scatter) of the period-luminosity relations, again because
of the reduced sensitivity of infrared luminosities to temperature
variations (across the instability strip), combined with the intrinsic
narrowness of the residual period-radius relations. And finally, (5)
at mid-infrared wavelengths, for the temperatures and surface
gravities encountered in Population I \& II Cepheids and RR Lyrae
stars, there are so few metallic line or molecular transitions in
those parts of the spectrum that atmospheric metallicity effects are
expected to have minimal impact on the calibration.\footnote{One
  significant exception at 4.5~$\mu$m has been noted for long-period
  Cepheids, where a CO molecular bandhead appears at temperatures
  below 4000K; see Marengo et al. 2010, Scowcroft et al. 2011, and
  Monson et al. 2012; however, the temperatures of the RR Lyrae
  variables, are so high in comparison to the long-period (classical)
  Cepheids, where this effect was discovered, that we expect no
  contribution of CO to the light or color curves of the RR Lyrae
  variables studied here. The W Virginis stars may be affected, and
  still need to be examined.} This is especially true for the RR~Lyrae
stars which are significantly hotter than their longer-period (cooler)
Cepheid counterparts.

As described in Freedman et al. (2012), the {\it Carnegie Hubble
  Program} (CHP) is designed to minimize and/or eliminate the
remaining known systematics in the measurement of the Hubble constant
using mid-infrared data from NASA's {\it Spitzer Space
  Telescope}. Here we broaden the base in two distinct ways: (a) the
incorporation of WISE mid-infrared data and (b) the preliminary
calibration of the Population II RR~Lyrae variables as mid-infrared
distance indicators. This new initiative is known as the {\it Carnegie
  RR~Lyrae Program} (CRRP).

\section{The Calibrators: WISE Observations}

WISE conducted an all-sky survey at four mid-infrared wavelengths,
3.4, 4.6, 12 and 22~$\mu$m (W1, W2, W3 and W4, hereafter).  As such all of
the RR~Lyrae variables having trigonometric parallaxes (Benedict et
al. 2011) were also observed by the satellite.  By design, the slowly
precessing orbit of WISE, allowed the satellite to scan across every
object on the sky at least 12 times (with progressively more coverage
at higher ecliptic latitudes). These successive observations were
obtained within a relatively narrow window of time (over about 18
hours for those fields nearest the ecliptic equator) with each
observation being separated by about 90 minutes (the orbital period of
the satellite). Fortuitously RR~Lyrae stars have periods that are
generally less than 16 hours, meaning that even the sparsest of these
multiple mid-infrared observations covered at least one full
pulsational cycle of these particular variable stars.

The light curves based on the time-resolved WISE observations of our
calibrating stars are shown in Figures 1 \& 2 for the four RR~Lyrae
variables, $SU ~Dra$, $RR ~Lyr$, $XZ ~Cyg$, \& $UV ~Oct$. These stars were
observed by WISE 51, 23, 24 and 23 times, respectively. Data were
retrieved from the WISE All-Sky Single Exposure (L1b) Source Table,
which is available at the Infrared Science Archive
(http://irsa.ipac.caltech.edu/Missions/wise.html).  The source
positions were queried with a 2.5 arcsec cone search radius, ignoring
observations flagged as contaminated by artifacts. The observations
are very uniformly distributed around the cycle and the resulting
light curves are exceedingly well defined. All three of the
shortest-wavelength light curves show convergence in their properties,
exemplified by their mutual phases, shapes and amplitudes\footnote{It
  should be noted that RR Lyr is saturated in the WISE data, and there
  is a documented flux over-estimation bias for saturated sources,
  especially in W2 (Cutri et al. 2012).  However, we have made no
  correction for the bias as it only becomes detectable for sources
  brighter than ~4.0 mag for W1 and ~6.0 mag for W2.}. As expected
these light curves closely track the anticipated light variations due
to surface-area variations of the star, where at these wavelengths the
sensitivity of the surface brightness to a temperature variation is
much diminished as compared to its sensitivity at optical
wavelengths. This then fully accounts for the mutual phasing (tracking
the radius variations, and not the off-set temperature variations),
the shape (the cycloid-like radius variation, in contrast to the
highly asymmetric color/temperature variation) and the low amplitude
(around 0.3~mag, peak-to-peak, in line with the small, radius-induced
cyclical change in surface area of these stars).

The non-parametric fitting methodology, GLOESS was used to derive
intensity-averaged magnitudes and amplitudes, as given in Table~1 (see
Persson et al. 2004 for a description and an early application of this
fitting technique). 

\section{RR~Lyrae Period-Luminosity Relations}

Table 1 contains the parameters needed to compute absolute mean
magnitudes. The parallaxes, E(B-V) reddenings, and Lutz-Kelker-Hanson
(LKH; Lutz and Kelker 1973; Hanson 1979) corrections as taken from
Benedict et al 2011, are listed for convenience.  We have converted
the A$_{V}$ extinctions listed by Benedict et al. (2011) to those in
the W1 and W2 bands using the Yuan et al. (2011) compilation of
A$_{WISE}$/E(B-V) results with A$_{V}$/E(B-V) = 3.1.  Indebetouw et
al. (2005) give values of A$_{Spitzer}$/$A_{K}$\footnote{The
  Indebetouw et al. values, strictly speaking, actually apply to the
  Spitzer channels 1 and 2 bands which are quite close to the W1 and
  W2 bands; the differences can be safely ignored.}, which we
converted to A$_{WISE}$/$A_{V}$ via the Cardelli et al.(1989) law. The
values of A$_{WISE}$/A$_{V}$ given by Yuan et al. (2011) and our
pseudo-values from Indebetouw et al. (2005) agree well.  Neither the
Yuan et al. (2011) nor Indebetouw et al. (2005) results extend to the
W3 band, and here we have referred to Fitzpatrick (1999) for an
approximate value.  The extinctions for the four stars are so small as
to make no difference to the absolute magnitude values and we take
$A_{W3}$/A$_{V}$ to be 0.01.  The adopted mid-IR extinctions
A$_{WISE}$/A$_{V}$ we adopt are 0.065, 0.052, and 0.01 for W1, W2, and
W3, respectively.  The adopted mid-IR extinctions A$_{WISE}$/E(B-V) are
  also given in Table 1. The above parameters and our observed mean
  magnitudes lead to the W1, W2, and W3 absolute magnitudes in Table
  1\footnote{Our magnitudes differ slightly from those in Benedict
    et. (2011). These differences stem from our use of different
    reddening law assumptions together with typographical errors in
    their Table 8 (Benedict 2013, private communication).}.  and the
  Period-Luminosity relations for RR~Lyrae variables follow:

$$ M_{[W1]} = -2.44 ~(\pm 0.95) \times log(P) -1.26 ~(\pm0.25)~~ \sigma = 0.10$$
$$ M_{[W2]} = -2.55 ~(\pm 0.89) \times log(P) -1.29 ~(\pm0.23)~~ \sigma = 0.10$$
$$ M_{[W3]} = -2.58 ~(\pm 0.97) \times log(P) -1.32 ~(\pm0.25)~~ \sigma = 0.10$$

The absolute magnitude values and the respective fits to the first two
WISE bands and as well as K-band data (from Benedict et al. 2011 and
Dall'Ora et al. 2004, respectively) are shown in Figure 3. Despite the
very small (less than a factor of two) range in period covered by
RR~Lyrae stars, the PL relations are well defined, largely because of
their intrinsically small scatter.  The intrinsic scatter is
especially well illustrated by the K-band PL relation, where we also
show the RR~Lyrae data of Dall'Ora et al. for 21 fundamental-mode
RR~Lyrae stars in the well-populated LMC globular cluster, Reticulum
(shifted by 18.47~mag). A comparison of these two datasets is very
illuminating. The slope of the adopted PL relation at K and the total
width of it, as defined by the two samples, are for all intents and
purposes, identical. The very good agreement in these two
independently-determined slopes and the small dispersion in each of
the datasets suggest that the means of the Milky Way variables are
already well constrained even though the Galactic calibrating sample
itself is currently very small.

On the other hand, we note that the small (observed) scatter of the
Milky Way RR~Lyrae variables around each of the adopted PL relations
is apparently at variance with the individually quoted error bars for
each of the calibrating variables. That is, the formal scatter of
$\pm0.10$~mag in the WISE PL relations is to be compared with the
quoted parallax errors on the individual distance moduli of $\pm0.22$,
$\pm0.16$, $\pm0.25$ and $\pm0.07$~mag for XZ~Cyg, UV~Oct, SU~Dra and
RR~Lyr, respectively. The average scatter for the variables
($\pm0.18$~mag) is then about two time larger that their observed scatter
around the PL fit. This suggests that the published errors may be
somewhat overestimated.  There are independent data that support this
assertion. The 10 Galactic Cepheids for which Benedict et al. (2010)
obtained parallaxes, using the same instrument, telescope and
reduction methodology have individually quoted internal errors in
their true distance moduli ranging from $\pm0.11$ to
$\pm0.30$~mag. Their average uncertainty is $\pm0.19$~mag, and yet,
once again, as with the RR~Lyrae variables the PL fit to these data
yields a formal dispersion of only $\pm0.10$~mag. In both cases the
observed dispersions, for the Galactic samples, are in total agreement
with independently determined dispersions for the much more robustly
determined dispersions for the LMC samples. We suggest therefore that
the random errors reported for the {\it HST} parallaxes for both the
Cepheids and for the RR~Lyrae variables may have been
over-estimated. This is not simply of academic interest. If the
observed scatter is used to calculate the systematic uncertainty in
the calibration of the RR~Lyrae PL relation that uncertainty would be
$0.10/\sqrt{4} = \pm0.05$~mag, a 2-3\% error in the Population~II
distance scale. However, if the quoted errors on the individual
distance moduli are used, then the uncertainty rises to $0.18/\sqrt{4}
= \pm0.09$~mag, a 5\% error. Similar conclusions would also apply to
the base uncertainty in the Cepheid distance scale using the Benedict
sample; is the uncertainty in the Galactic Cepheid zero point 1.6\% in
distance, or is it 3.0\%?  It is therefore important to note that in
their first paper discussing the use of {\it FGS} on {\it HST},
Benedict et al. (2002) state that the ``standard deviations of the
{\it HST} and {\it Hipparcos} data points may have been overstated by
a factor of $\sim$1.5.''  and since the {\it Hipparcos} errors had
been subjected to many confirming tests ``... that it is likely that
the {\it HST} errors are overstated.''  Parallaxes from Gaia are
anxiously awaited; they will improve the number of calibrators by
orders of magnitude and convincingly set the zero point.

In Figure 3 we show, using thick vertical lines, the full magnitude of
the LKH corrections as published by Benedict et al (2011) and applied
to the true distance moduli used here.  It is noteworthy that, if
these corrections had not been applied, the dispersion in the data
points around the fit would have exceeded the independently determined
dispersion from the Reticulum data, and the slope of the Milky Way
solution would have been more shallow than the LMC slope. We take the
final agreement of both the slopes and the dispersions to suggest that
the individually determined and independently-applied LKH corrections
are appropriate.

Finally, it needs to be noted that Klein et al. (2011) have published
slopes that are much shallower than the ones derived here (e.g., -1.7
compared to our -2.6). This is because in their Bayesian analysis they
chose to leave the overtone pulsators in the global solution, without
correcting them to their equivalent fundamental periods. We have
recomputed the slopes from their data after applying the appropriate
period shift to the overtones, and those PL slopes are plotted in
Figure 4. Their slopes and ours now agree well within the errors, but
they are still systematically somewhat shallower than our solutions.

\section{The Run of PL Slope with Wavelength}

For Cepheids it is well known that the slope of the PL relation is a
monotonically increasing function of wavelength. In Figure 4 we show
that the same overall trend is now made explicit for the first time
for the RR~Lyrae variables as well, and for the same physical
reasons. As one moves from shorter to longer wavelengths one is moving
from PL relations where the slope is dominated by the trend of
decreasing temperature (i.e., decreasing surface brightness) with
period, to relations that are dominated by the opposing run of
increasing mean radius with period. The plotted slopes of the optical
and near-infrared PL relations are representative of a variety of
published studies (e.g, Catelan, Pritzl \& Smith 2004, Benedict et
al. 2011, Dall'Ora et al. 2004) while the mid-IR slopes are from this
study. As the relative contribution from the temperature-sensitive
surface brightness drops off with wavelength, the observed slope is
expected to asymptotically approach the wavelength-independent
(geometric) slope of the Period-Area relation. That behavior is indeed
seen in Figure 4. Moreover the level at which the plateau is occurring
would suggest that the period-radius relation of Burki \& Meylan
(1986) (giving a slope of -2.60, based on Baade-Wesselink studies) is
marginally preferred over the period-radius (slope = -3.25) and
period-radius-metallicity (slope = -2.90) solutions given by Marconi
et al. (2005)\footnote{The referee has correctly pointed out that
  ``the Period-Radius relation provided by Burki \& Meylan (1986) is
  based on a mix of $\delta$ Scuti, RR~Lyrae and Type~II (W Virginis)
  stars'', and that ``there is is no solid reason why the quoted
  pulsators should obey the same PR relation.'' At the same
  time, he/she notes that ``the Period-Radius relation provided by
  Marconi et al. (2005) is based on a set of RR~Lyrae models that
  cover more than two dex in metal abundance ... and they also account
  for, at fixed metal abundance, possible evolutionary effects.''}

From a practical point of view it is not immediately clear what
advantage the increased slope of the long-wavelength PL relations
would have to offer applications to the distance scale, until it is
realized that increased slope in the PL relation is causally and
physically connected to decreased width (i.e., decreased intrinsic
scatter and therefore increased precision) in the PL relation as
proven in the general case by Madore \& Freedman (2012). This effect
can be seen for the RR~Lyrae variables in Figure 2 of Catelan, Pritzl
\& Smith (2004), and it is apparent here in Figure 3 where the scatter
has already reached a minimum in the K-band where simultaneously the
plateau in slope (seen in Figure 4) is very nearly complete.

\section{A First Test of the Metallicity Dependence in the Mid-IR}

In Figure 5 we plot the measured magnitude residuals from the
[W1] 3.4~$\mu$m PL relation versus the published metallicities of the four
RR~Lyrae stars in our sample, as given in Table 1 of Benedict et
al. (2011). The RR~Lyrae stars only sample a 0.4~dex range in [Fe/H]
so the test is not a strong one, but there is clearly no significant
dependence of the already small magnitude residuals on metallicity.

\section{Conclusions}
As can be dramatically seen in the study of Catelan, Pritzl \& Smith
(2008, especially their Figure 2) operating anywhere in the near to
mid-infrared, from H = 1.6~$\mu$m (accessible to {\it HST}) to 3.6~$\mu$m
(accessible to Spitzer now, and with JWST in the near future) will
each accrue the benefits of low scatter and ever decreasing
sensitivity (with wavelength) to line-of-sight extinction.  Collecting
power, availability and spatial resolution will determine which of
these instruments will be used at any given time. But suffice it to
say that the Population~II RR~Lyrae variables are proving themselves
to be a powerful means of establishing an independent, highly precise
and accurate distance scale that is completely decoupled in its
systematics from the Population~I Cepheid path to the extragalactic
distance scale and the Hubble constant.

\medskip
\medskip
This work is based in part on observations made with the {\it
  Wide-field Infrared Survey Explorer} (WISE), which was is operated
by the Jet Propulsion Laboratory, California Institute of Technology
under a contract with NASA.  Support for this work was provided by
NASA through an award issued by JPL/Caltech. This research also made
use of the NASA/IPAC Extragalactic Database (NED) and the NASA/ IPAC
Infrared Science Archive (IRSA), both of which are operated by the Jet
Propulsion Laboratory, California Institute of Technology, under
contract with the National Aeronautics and Space Administration.  We
thank Fritz Benedict for numerous frank and useful communications.
The referee was especially helpful in bringing this paper to a more
correct and fruitful completion.

\vfill\eject

\centerline{\bf References \rm}
\vskip 0.1cm
\vskip 0.1cm

\par\noindent Benedict, G. F., McArthur, B.E., Feast, M.W., et al.  2011, \aj, 142, 187

\par\noindent Benedict, G. F., McArthur, B.E., Fredrick, L.W., et al.  2002, \aj, 124, 1695

\par\noindent Burki, G., \& Meylan, G. 1986, \aap, 159, 261

\par\noindent Cardelli, J.A., Clayton, G.C., \& Mathis, J.S 1989, \apj, 345, 245 

\par\noindent Catelan, M., Pritzl, B.J., \& Smith, H.A. 2004, \apjs, 154, 633

\par\noindent Cutri et al. 2012, WISE All-Sky Release Explanatory Supplement, 
\par http://wise2.ipac.caltech.edu/docs/release/allsky/expsup/

\par\noindent Dall'Ora, M., Storm, J., Bono, G., et al.  2004, \apj, 610, 269

\par\noindent Fitzpatrick, E.L. 1999, \pasp, 111, 63 

\par\noindent Freedman, W.L., \& Madore, B.F. 2010, ARA\&A, 48, 673

\par\noindent Indebetouw, R., Mathis, J. S., \& Babler, B. L. et al. 2005, \apj, 619, 931

\par\noindent Klein, C.R., Richards, J.W., Butler, N.R., \& Bloom, J.S. 2011, \apj, 738, 185

\par\noindent Madore, B.F. \& Freedman, W.L. 2010, \apj, 744, 132

\par\noindent Marconi, M., Nordgren, T., Bono, G., Schnider, G., \& Caputo, F. 2005, \apjl, 623, 133

\par\noindent McGonegal, R., McAlary, C.W., McLaren, R.A. \& Madore, B.F. 1983, \apjl, 257, 33

\par\noindent Marengo, M., Evans, N.R., Barmby, P., et al. 2010, \apj, 709, 120

\par\noindent Monson, A., Freedman, W.L., Madore, B.F., et al. 2012, \apj, 759, 146

\par\noindent Persson, S.E., Madore, B.F., Krzeminski, W., et al. 2012, \aj, 128, 2239

\par\noindent Scowcroft, V., Freedman, W.L., Madore, B.F.,  et al.  2011, \apj, 743, 76

\par\noindent Sollima, A., Cacciari, C., \& Valenti, E.  2006, \mnras, 372, 1675

\par\noindent Wright, E.L., Eisenhardt, P.R.M., Mainzer, A.K., et al.  2010, AJ, 140, 1868

\par\noindent Yuan,H.B., Liu, X.W., \& Xiang, M.S. 2013, \mnras, 430, 2188

\vfill\eject

\begin{figure*}
    \centering
  \includegraphics[width=12.0cm, angle=-0]{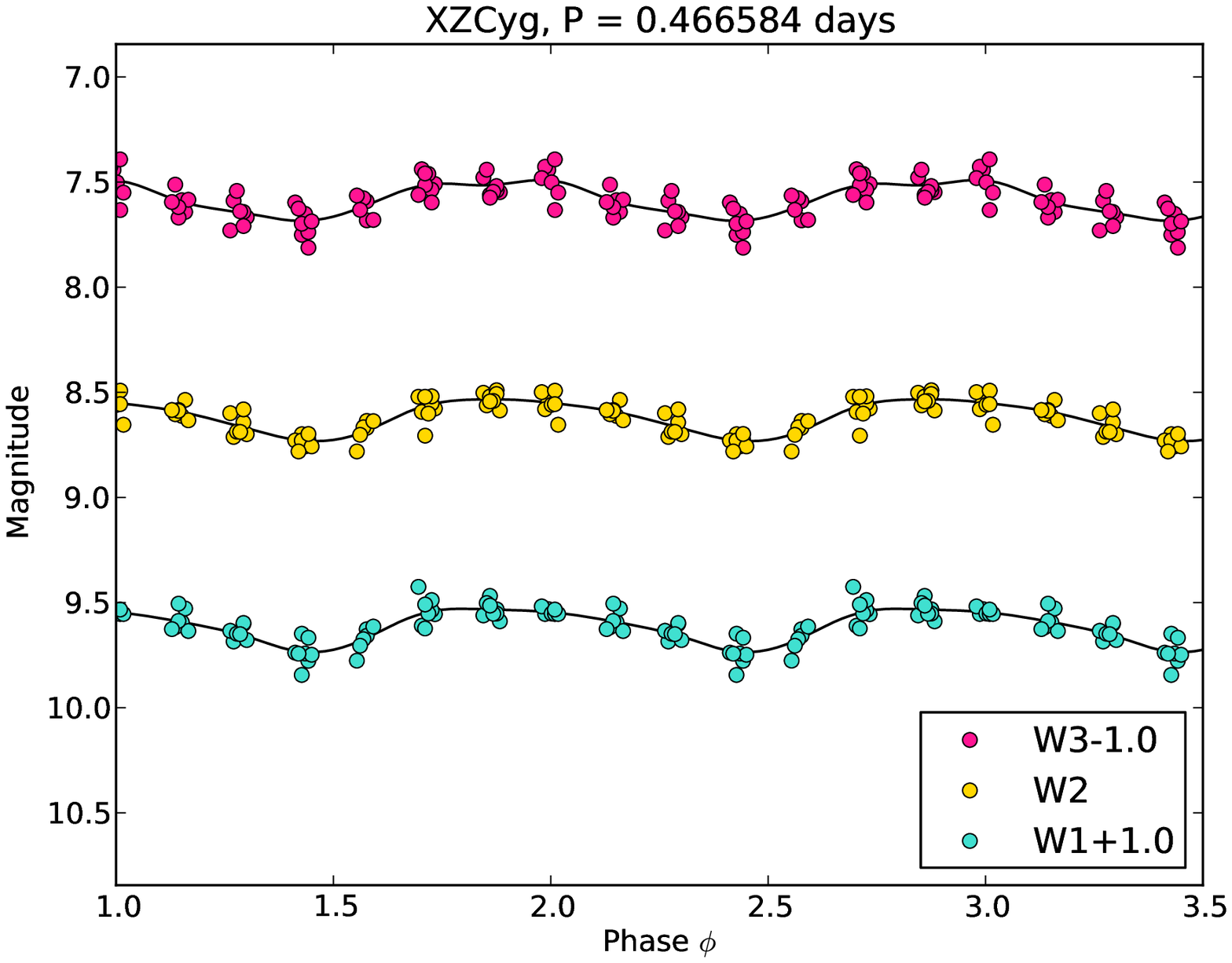}
  \includegraphics[width=12.0cm, angle=-0]{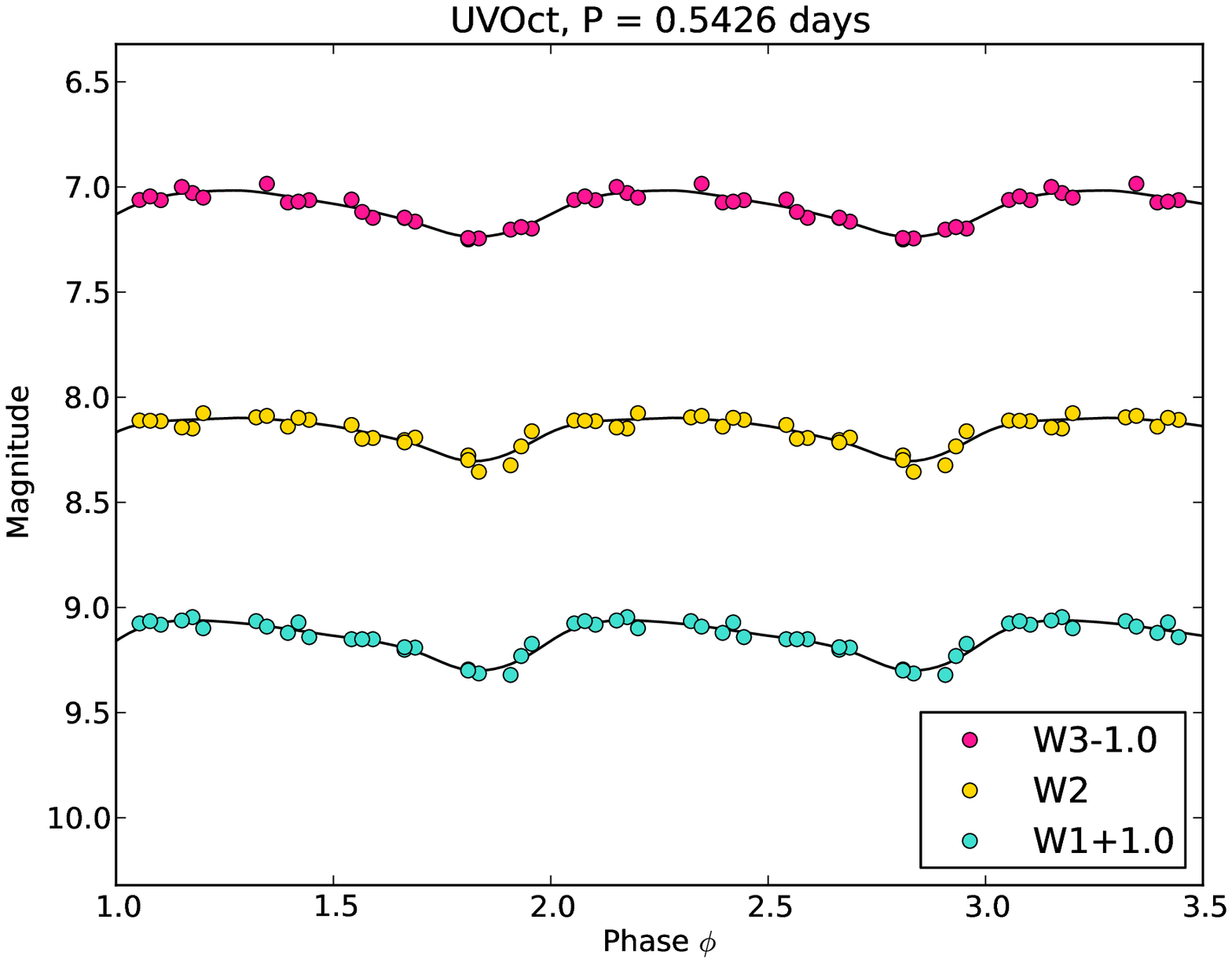}
\caption{WISE Mid-Infrared light curves for XZ~Cyg (upper panel) and
  UV~Oct (lower panel) phase-folded over two and a half cycles using
  the periods given in the titles. GLOESS fits are shown as solid
  black lines.}
\end{figure*}

\begin{figure*}
    \centering
  \includegraphics[width=12.0cm, angle=-0]{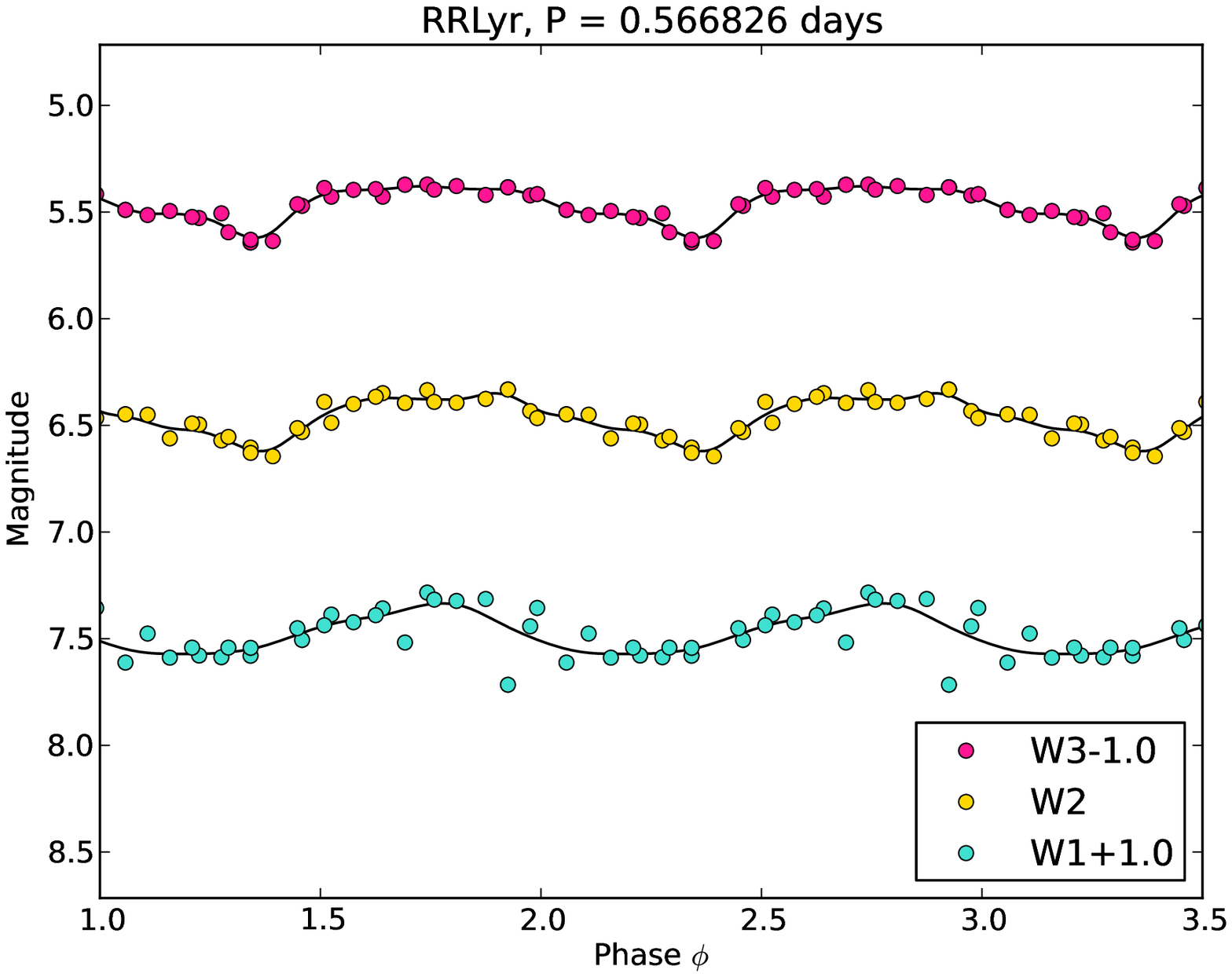}
  \includegraphics[width=12.0cm, angle=-0]{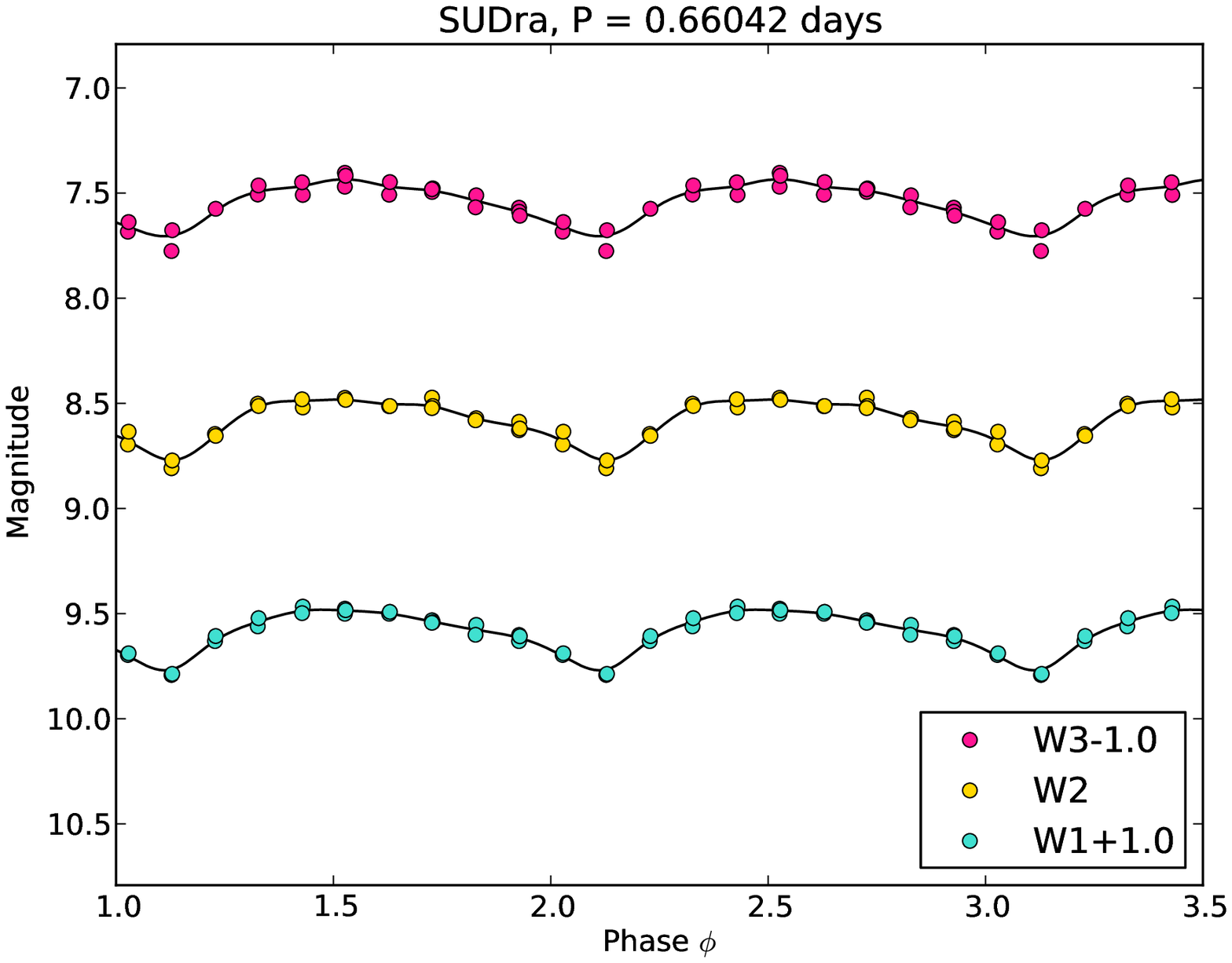}
\caption{WISE Mid-Infrared light curves for RR~Lyr (upper panel) and
  SU~Dra (lower panel) phase-folded over two and a half cycles using
  the periods given in the titles. GLOESS fits are shown as solid
  black lines.}
\end{figure*}

\begin{figure}
\includegraphics [width=19.0cm, angle=270.] {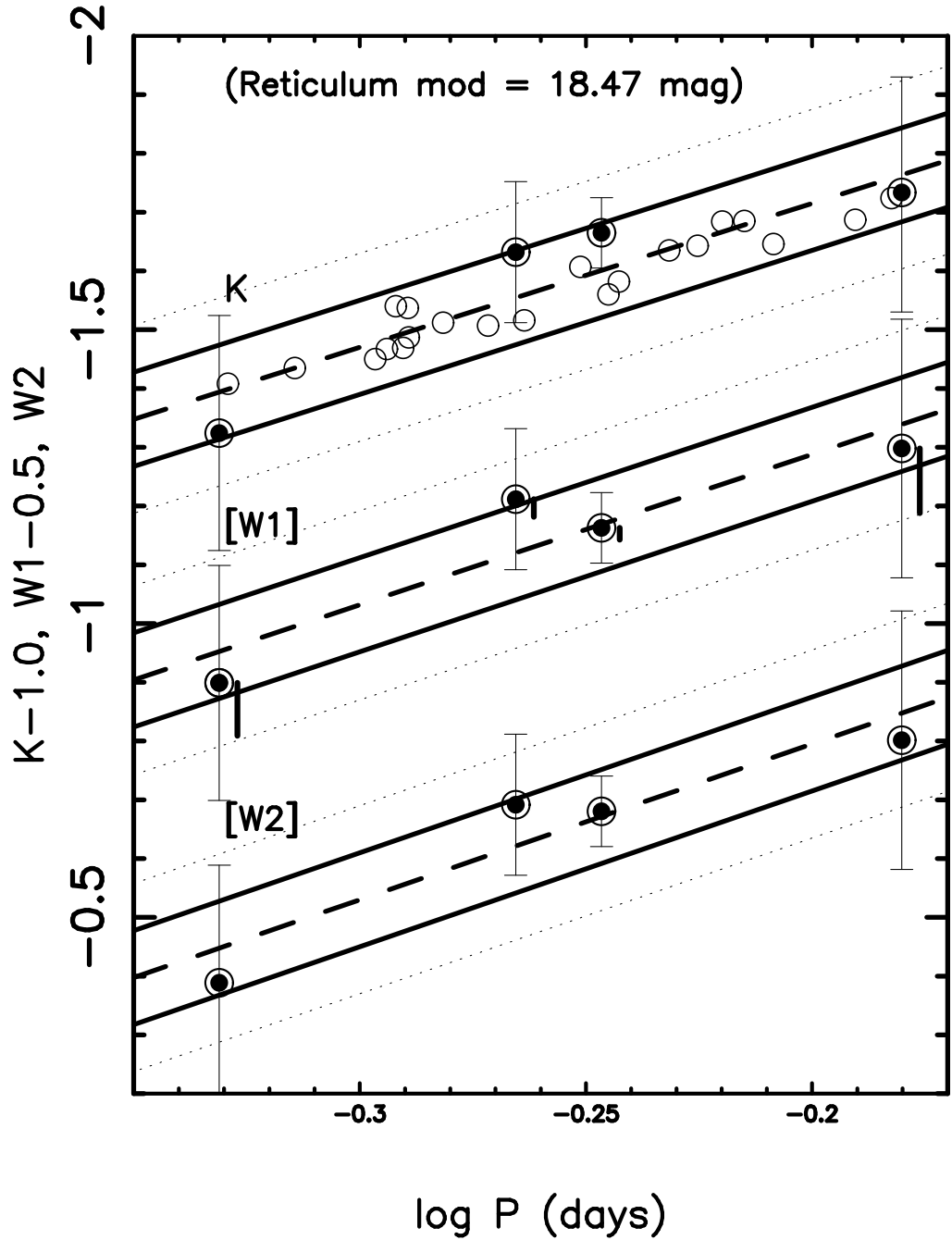}
\caption{\footnotesize RR~Lyrae PL relations in the K-band (top) and
  the two WISE bands [W1] (middle and [W2] bottom.  The K-band
  relation also contains data from Dall'Ora et al. (2004) for the LMC
  globular cluster, Reticulum, shifted by 18.47~mag. (This distance
  modulus shift is remarkably close to the independently-determined
  true modulus of 18.48~mag recently reported by Monson et al. 2012
  for the LMC Cepheid mid-infrared distance modulus.) The Reticulum
  data are shown only for the RRab (fundamental) pulsators, and are
  presented here to illustrate that they are consistent in slope and
  scatter in comparison with the Galactic calibration. A detailed
  discussion of Reticulum will be given in a forthcoming paper (Monson
  et al. 2013). The solid lines flanking each of the fitted PL
  relations are each separated by two sigma from their respective
  ridge lines. Despite the small numbers of stars represented here the
  full width of the PL relation in each of the bands is well
  defined. The solid vertical lines to the right of each of the [W1]
  data points represents the LKH correction applied by
  Benedict et al. (2011). }
\end{figure}

\begin{figure*}
    \centering
  \includegraphics[width=18.0cm, angle=-0]{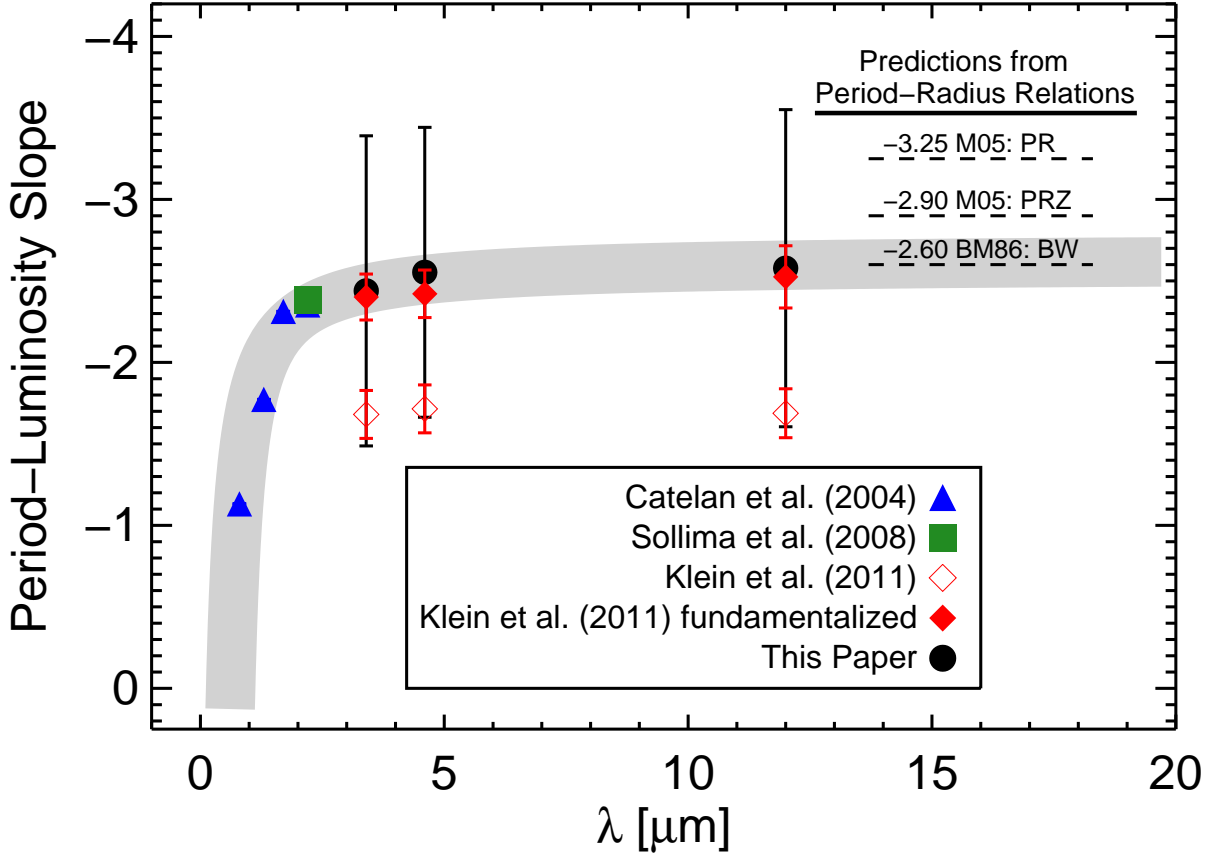}
\caption{\footnotesize The expected monotonic increase of the slope of
  the RR~Lyrae Period-Luminosity relation as a function of increasing
  wavelength. The asymptotic behavior of the slope, approaching a
  value of about -2.6 indicates that the PL relation is converging on
  the Period-Radius relation, as theory would predict, given that the
  sensitivity of the surface brightness to temperature rapidly drops
  as one progressively moves into the infrared. The open diamonds are
  the slopes published by Klein et al. (2011); the filled (red)
  diamonds indicate the ``fundamentalized'' slopes (where we have
  corrected the periods of the overtone pulsators to their
  corresponding fundamental periods by adding 0.127 to the log of
  their observed periods, as in Dall'Ora et al. 2004), based on the
  data published by Klein et al. (2011) and re-fit for this paper. The
  optical and near-infrared PL relation slopes are from Catelan,
  Pritzl \& Smith (2004), Benedict et al. (2011) and Sollima, Cacciari
  \& Valenti al. (2006), while the mid-IR slopes are from this
  study. The equivalent slopes derived from Period-Radius relations
  are from Burki \& Meylan (1986; BM86) and Marconi et al. (2005;
  M05).}
\end{figure*}

\begin{figure*}
    \centering
  \includegraphics[width=6.0cm, angle=-90]{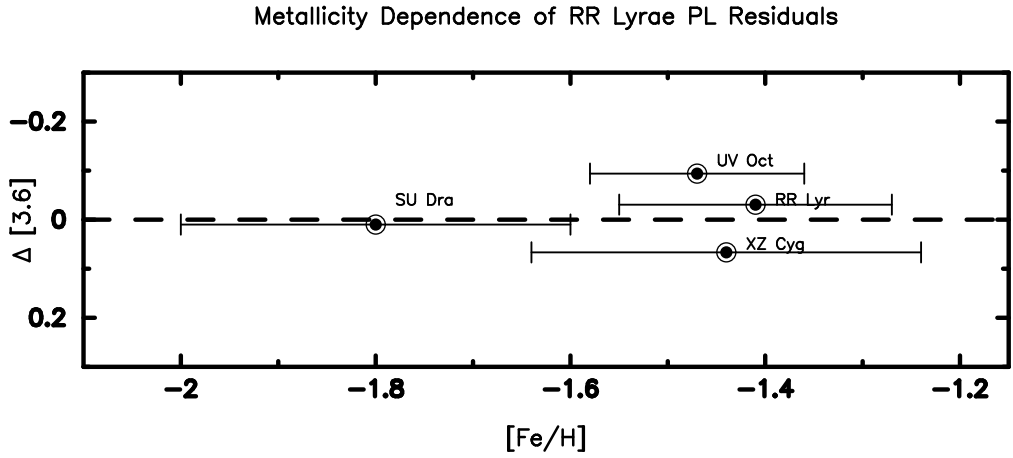}
\caption{Mid-Infrared [W1] (3.4~$\mu$m) deviations from the mean
  Period-Luminosity relation as a function of metallicity. The
  currently available sample is small, and the metallicity range is
  limited. No obvious correlation is seen.}
\end{figure*}

\clearpage
\begin{deluxetable}{lcccc}
\tablecolumns{5}
\tablewidth{5.1truein}
\tablecaption{Mid-Infrared (WISE) Magnitudes for Galactic RR~Lyrae Variables}
\tablehead{
\colhead{Name}  & \colhead{XZ~Cyg}  & \colhead{UV~Oct} & \colhead{RR~Lyr}  & \colhead{SU~Dra}
}
\startdata
log(P) & -0.33107 & -0.26552 & -0.24655 & -0.18018 \\
parallax (mas)  & 1.67$\pm$ 0.17 & 1.71$\pm$ 0.10 & 3.77$\pm$ 0.13 & 1.42$\pm$ 0.16 \\
A$_V$\tablenotemark{a} & ~0.30 & ~0.28 & ~0.13 & ~0.03 \\
A$_{W1}$ & ~0.020 & ~0.018 & ~0.008 & ~0.002 \\
A$_{W2}$ & ~0.016 & ~0.015 & ~0.007 & ~0.002 \\
A$_{W3}$ & ~0.002 & ~0.001 & ~0.001 & ~0.000 \\
LKH Corr. & -0.09 & -0.03 & -0.02 & -0.11 \\
$(m-M)_o$ & ~8.98 & ~8.87 & ~7.14 & ~9.35 \\
 &  &  &  &  \\
m[W1]  & 8.610 & 8.156 & 6.469 & 8.584 \\
$\sigma$[W1]  & 0.010 & 0.017 & 0.017 & 0.019 \\
Ampl [W1]  & 0.21~ & 0.24~ & 0.24~ & 0.29~ \\
 &  &  &  &  \\
m[W2]  & 8.616 & 8.172 & 6.456 & 8.580 \\
$\sigma$[W2]  & 0.010 & 0.014 & 0.017 & 0.019 \\
Ampl [W2]  & 0.20~ & 0.21~ & 0.27~ & 0.29~ \\
 &  &  &  &  \\
m[W3]  & 8.579 & 8.110 & 6.460 & 8.543 \\
$\sigma$[W3]  & 0.010 & 0.016 & 0.015 & 0.018 \\
Ampl [W3]  & 0.19~ & 0.22~ & 0.24~ & 0.27~ \\
 &  &  &  &  \\
M[W1]& -0.39 (0.23) & -0.73 (0.13) &  -0.68 (0.08) &  -0.77 (0.26) \\
M[W2]& -0.38 (0.23) & -0.71 (0.13) &  -0.69 (0.08) &  -0.77 (0.26) \\
M[W3]& -0.40 (0.23) & -0.76 (0.13) &  -0.68 (0.08) &  -0.81 (0.26) \\
\enddata
\tablenotetext{a}{ \ \ Benedict et al. (2011), where $A_{band}$/E(B-V) = 0.20, 0.16, 0.03 for W1, W2 and W3, respectively.}
\end{deluxetable}

\vfill\eject


 \begin{deluxetable}{rrlrlrl} 
 \tablecolumns{7} 
 \tablewidth{0pc} 
 \tablecaption{WISE Observations of RR Lyrae} 
 \tablehead{ 
 \colhead{MJD(2,400,000.+)} & \colhead{3.4~$\mu$m}& \colhead{err} &\colhead{4.6~$\mu$m}&\colhead{err}&\colhead{12~$\mu$m} &\colhead{err}}
 \startdata
55312.047509&6.612&0.019&6.448&0.022&6.490&0.014\\
55312.179813&6.542&0.034&6.554&0.019&6.595&0.019\\
55312.312118&6.387&0.031&6.488&0.020&6.428&0.013\\
55312.378206&6.358&0.049&6.349&0.021&6.428&0.019\\
55312.444422&6.317&0.030&6.390&0.019&6.395&0.014\\
55312.510510&6.314&0.032&6.376&0.018&6.420&0.019\\
55312.576726&6.356&0.034&6.466&0.017&6.416&0.016\\
55312.642814&6.476&0.031&6.450&0.021&6.514&0.016\\
55312.709030&6.579&0.026&6.496&0.021&6.528&0.016\\
55312.775119&6.580&0.041&6.604&0.019&6.643&0.017\\
55312.775246&6.543&0.029&6.629&0.020&6.630&0.020\\
55312.841334&6.506&0.028&6.531&0.019&6.471&0.018\\
55312.907423&6.423&0.032&6.400&0.016&6.396&0.013\\
55312.973639&6.518&0.04 &6.395&0.018&6.372&0.017\\
55313.039855&6.323&0.037&6.394&0.020&6.378&0.018\\
55313.105943&6.716&0.04 &6.332&0.017&6.384&0.019\\
55313.238247&6.589&0.030&6.561&0.017&6.495&0.016\\
55313.304463&6.587&0.029&6.571&0.015&6.506&0.016\\
55313.436767&6.437&0.044&6.390&0.018&6.387&0.014\\
55313.502856&6.390&0.037&6.366&0.013&6.392&0.014\\
55313.569072&6.284&0.034&6.335&0.020&6.371&0.016\\
55313.701376&6.442&0.030&6.433&0.028&6.422&0.014\\
55313.833680&6.542&0.032&6.491&0.022&6.523&0.017\\
55321.904496&6.451&0.038&6.513&0.018&6.463&0.016\\
 \enddata
 \end{deluxetable}

\clearpage

 \begin{deluxetable}{rrlrlrl} 
 \tablecolumns{7} 
 \tablewidth{0pc} 
 \tablecaption{WISE Observations of SU Draconis} 
 \tablehead{ 
 \colhead{MJD(2,400,000.+)} & \colhead{3.4~$\mu$m}& \colhead{err} &\colhead{4.6~$\mu$m}&\colhead{err}&\colhead{12~$\mu$m} &\colhead{err}}
 \startdata
55315.013213&8.561&0.019&8.502&0.017&8.507&0.037\\
55315.145517&8.501&0.021&8.474&0.022&8.470&0.043\\
55315.145644&8.477&0.022&8.479&0.021&8.404&0.039\\
55315.277821&8.544&0.023&8.523&0.019&8.482&0.040\\
55315.277949&8.532&0.022&8.473&0.023&8.495&0.040\\
55315.344037&8.601&0.017&8.581&0.013&8.568&0.040\\
55315.410125&8.631&0.022&8.628&0.019&8.570&0.044\\
55315.410253&8.602&0.021&8.588&0.017&8.589&0.046\\
55315.476341&8.697&0.020&8.696&0.017&8.684&0.046\\
55315.542559&8.792&0.021&8.809&0.017&8.776&0.052\\
55315.608647&8.631&0.017&8.646&0.019&\nodata&\nodata\\
55315.674863&8.522&0.021&8.513&0.016&8.464&0.043\\
55315.740952&8.498&0.017&8.481&0.017&8.449&0.043\\
55315.807167&8.484&0.025&8.484&0.020&8.418&0.037\\
55315.873256&8.501&0.020&8.514&0.019&8.508&0.039\\
55315.939472&8.541&0.020&8.513&0.020&8.478&0.040\\
55316.005560&8.554&0.018&8.571&0.016&8.511&0.040\\
55316.071776&8.608&0.021&8.620&0.019&8.608&0.045\\
55316.137864&8.689&0.021&8.635&0.019&8.638&0.043\\
55316.204080&8.787&0.018&8.772&0.024&8.677&0.043\\
55316.270169&8.607&0.016&8.655&0.018&8.575&0.042\\
55316.402473&8.467&0.019&8.520&0.022&8.509&0.045\\
55316.534777&8.492&0.026&8.513&0.020&8.448&0.039\\
 \enddata
 \end{deluxetable}

\clearpage

 \begin{deluxetable}{rrlrlrl} 
 \tablecolumns{7} 
 \tablewidth{0pc} 
 \tablecaption{WISE Observations of UV Oct} 
 \tablehead{ 
 \colhead{MJD(2,400,000.+)} & \colhead{3.4~$\mu$m}& \colhead{err} &\colhead{4.6~$\mu$m}&\colhead{err}&\colhead{12~$\mu$m} &\colhead{err}}
 \startdata
55270.919633&8.082&0.025&8.114&0.020&8.063&0.031\\
55271.052065&8.091&0.021&8.089&0.015&7.985&0.029\\
55271.184369&8.151&0.019&8.194&0.020&8.147&0.032\\
55271.316673&8.314&0.016&8.355&0.018&8.245&0.032\\
55271.382761&8.173&0.028&8.162&0.019&8.198&0.032\\
55271.448977&8.065&0.019&8.112&0.017&8.045&0.030\\
55271.515192&8.099&0.020&8.076&0.016&8.051&0.029\\
55271.581281&8.065&0.018&8.096&0.016&\nodata&\nodata\\
55271.647496&8.141&0.019&8.108&0.021&8.063&0.030\\
55271.713585&8.151&0.022&8.198&0.019&8.119&0.034\\
55271.779800&8.191&0.020&8.192&0.022&8.165&0.030\\
55271.845888&8.300&0.022&8.299&0.020&8.243&0.034\\
55271.846016&8.295&0.022&8.277&0.022&8.250&0.032\\
55271.912104&8.231&0.019&8.234&0.017&8.191&0.032\\
55271.978320&8.076&0.026&8.111&0.016&8.062&0.028\\
55272.044408&8.046&0.018&8.149&0.017&8.029&0.029\\
55272.176712&8.071&0.021&8.098&0.019&8.070&0.033\\
55272.242928&8.151&0.017&8.132&0.013&8.060&0.029\\
55272.309016&8.201&0.022&8.204&0.019&8.148&0.033\\
55272.309144&8.189&0.022&8.215&0.022&8.146&0.031\\
55272.441448&8.321&0.021&8.324&0.020&8.203&0.033\\
55272.573751&8.062&0.023&8.144&0.028&8.000&0.030\\
55272.706056&8.121&0.017&8.140&0.013&8.074&0.029\\
 \enddata
 \end{deluxetable}

 \begin{deluxetable}{rrlrlrl} 
 \tablecolumns{7} 
 \tablewidth{0pc} 
 \tablecaption{WISE Observations of XZ Cyg} 
 \tablehead{ 
 \colhead{MJD(2,400,000.+)} & \colhead{3.4~$\mu$m}& \colhead{err} &\colhead{4.6~$\mu$m}&\colhead{err}&\colhead{12~$\mu$m} &\colhead{err}}
 \startdata
55334.008212&8.589&0.016&8.586&0.017&8.548&0.037\\
55334.140516&8.635&0.016&8.633&0.017&8.584&0.040\\
55334.272820&8.748&0.019&8.757&0.017&8.687&0.048\\
55334.339035&8.614&0.018&8.637&0.021&8.680&0.049\\
55334.405124&8.555&0.019&8.577&0.018&8.509&0.041\\
55334.471212&8.552&0.028&8.508&0.020&8.519&0.039\\
55334.471339&8.532&0.018&8.490&0.021&8.540&0.048\\
55334.537428&8.554&0.019&8.654&0.020&8.550&0.040\\
55334.603643&8.528&0.025&8.536&0.015&8.642&0.040\\
55334.669732&8.678&0.020&8.699&0.020&8.667&0.046\\
55334.735820&8.777&0.023&8.725&0.021&8.738&0.044\\
55334.735947&8.667&0.023&8.698&0.015&8.812&0.046\\
55334.868124&8.536&0.023&8.553&0.022&8.536&0.038\\
55334.868251&8.488&0.026&8.519&0.015&8.597&0.045\\
55334.934340&8.553&0.019&8.541&0.017&8.546&0.044\\
55335.000428&8.533&0.027&8.556&0.020&8.392&0.035\\
55335.000555&8.554&0.022&8.492&0.020&8.633&0.046\\
55335.066644&8.593&0.019&8.612&0.018&8.586&0.047\\
55335.132732&8.602&0.026&8.643&0.020&8.642&0.041\\
55335.132859&8.597&0.019&8.581&0.018&8.709&0.047\\
55335.198947&8.742&0.017&8.757&0.021&8.651&0.044\\
55335.265036&8.627&0.021&8.669&0.016&8.591&0.039\\
55335.265163&8.658&0.023&8.636&0.021&8.682&0.044\\
55335.331251&8.552&0.017&8.601&0.021&8.461&0.041\\
55335.397340&8.468&0.019&8.520&0.016&8.560&0.039\\
55335.397467&8.514&0.024&8.543&0.020&8.574&0.044\\
\tablebreak
55335.463555&8.552&0.017&8.559&0.019&8.501&0.043\\
55335.529644&8.587&0.019&8.588&0.016&8.619&0.040\\
55335.529771&8.505&0.021&8.584&0.021&8.669&0.045\\
55335.595859&8.650&0.017&8.688&0.020&8.639&0.047\\
55335.661948&8.844&0.022&8.729&0.019&8.698&0.046\\
55335.662075&8.648&0.021&8.699&0.023&8.751&0.050\\
55335.728163&8.674&0.017&8.666&0.019&8.577&0.044\\
55335.794252&8.623&0.023&8.521&0.019&8.459&0.040\\
55335.794379&8.509&0.023&8.706&0.029&8.514&0.043\\
55335.860467&8.502&0.018&8.562&0.017&8.441&0.040\\
55335.926556&8.531&0.021&8.546&0.020&8.442&0.040\\
55335.992771&8.617&0.019&8.604&0.017&8.512&0.040\\
55336.058860&8.648&0.021&8.687&0.020&8.542&0.042\\
55336.125075&8.743&0.020&8.781&0.026&8.626&0.041\\
55336.191164&8.705&0.022&8.702&0.020&8.632&0.045\\
55336.257379&8.610&0.019&8.593&0.024&8.439&0.036\\
55336.323468&8.561&0.021&8.502&0.019&8.478&0.039\\
55336.389683&8.554&0.024&8.580&0.026&8.427&0.038\\
55336.455772&8.626&0.022&8.584&0.017&8.595&0.048\\
55336.521985&8.685&0.017&8.712&0.014&8.589&0.039\\
55336.588074&8.738&0.021&8.729&0.023&8.597&0.044\\
55336.654289&8.776&0.018&8.781&0.014&8.564&0.040\\
55336.720377&8.425&0.021&8.521&0.017&8.561&0.039\\
55336.852681&8.518&0.018&8.499&0.016&8.481&0.038\\
55336.984985&8.634&0.019&8.599&0.017&8.730&0.046\\
 \enddata
 \end{deluxetable}

 \end{document}